\documentclass[journal]{IEEEtran}
\usepackage{graphicx}
\usepackage{amsmath}

\usepackage{amssymb}
\usepackage{bm}
\usepackage{float}
\usepackage{booktabs}
\usepackage{multirow}
\usepackage{makecell}
\usepackage{verbatim}
\usepackage{enumitem}
\usepackage{soul}

\usepackage{csquotes}

\usepackage{booktabs}
\usepackage{multirow}
\usepackage{makecell}
\usepackage{lipsum} 
\usepackage{multirow}

\usepackage{diagbox}
\usepackage{bm}
\usepackage{booktabs}
\usepackage{siunitx} 
\usepackage[T1]{fontenc}

\usepackage{acronym}
\usepackage[acronym,shortcuts]{glossaries}
\usepackage{cite}
\usepackage{amsmath}
\makeglossaries

\usepackage{algorithm}
\usepackage{algorithmic}
\floatname{algorithm}{}
\usepackage{threeparttable}

\makeatletter
\newcount\SOUL@minus 
\makeatother

\begin{document}
	\title{RainGaugeNet: CSI-Based Sub-6 GHz Rainfall Attenuation Measurement and Classification for ISAC Applications}
	\author{~Yan~Li,~Jie~Yang,~\IEEEmembership{Member,~IEEE},~Yixuan~Huang,~Tao~Yang,~Chao-Kai~Wen,~\IEEEmembership{Fellow,~IEEE}
		and~Shi~Jin,~\IEEEmembership{Fellow,~IEEE}

		\thanks{
			{Y.~Li} and {Y.~Huang} are with the National Mobile Communications Research
			Laboratory, Southeast University, Nanjing 210096, P. R. China, Email:
			{\rm  leeyan@seu.edu.cn} and {\rm  huangyx@seu.edu.cn}.
			{J.~Yang} is with Frontiers Science Center for Mobile Information Communication and Security and Key Laboratory of Measurement and Control of Complex Systems of Engineering, Ministry of Education, Southeast University, Nanjing 210096, P. R. China, Email:{\rm  yangjie@seu.edu.cn}.
			{T.~Yang} is with the National Engineering Research Center of  Water Resources Efficient Utilization and Engineering Safety, Institute of Water Science and Technology, Hohai University, Nanjing, 210098, China, Email:  {\rm  tao.yang@hhu.edu.cn}.
			{C.-K.~Wen} is with the Institute of Communications Engineering, National Sun Yat-sen University, Kaohsiung 80424, Taiwan, Email:  {\rm chaokai.wen@mail.nsysu.edu.tw}.
			{S.~Jin} is with the National Mobile Communications Research Laboratory, Southeast University and Frontiers Science Center for Mobile Information Communication and Security, Nanjing 210096, P. R. China, Email: {\rm  jinshi@seu.edu.cn}.}
	}		
	\maketitle
	\begin{abstract}
		Rainfall impacts daily activities and can lead to severe hazards such as flooding. Traditional rainfall measurement systems often lack granularity or require extensive infrastructure. While the attenuation of electromagnetic waves due to rainfall is well-documented for frequencies above 10 GHz, sub-6 GHz bands are typically assumed to experience negligible effects. However, recent studies suggest measurable attenuation even at these lower frequencies. This study presents the first channel state information (CSI)-based measurement and analysis of rainfall attenuation at 2.8 GHz. The results confirm the presence of rain-induced attenuation at this frequency, although classification remains challenging. The attenuation follows a power-law decay model, with the rate of attenuation decreasing as rainfall intensity increases. Additionally, rainfall onset significantly increases the delay spread. Building on these insights, we propose RainGaugeNet, the first CSI-based rainfall classification model that leverages multipath and temporal features. Using only 20 seconds of CSI data, RainGaugeNet achieved over 90\% classification accuracy in line-of-sight scenarios and over 85\% in non-line-of-sight scenarios, significantly outperforming state-of-the-art methods. 
	\end{abstract}
	\IEEEpeerreviewmaketitle

	\section{Introduction}

	Integrated sensing and communications (ISAC) has emerged as a transformative paradigm, integrating sensing capabilities into future mobile networks to create perceptive systems capable of ubiquitously monitoring the environment \cite{liufan}. Among its many applications, weather monitoring is a crucial component of ISAC \cite{li2023integrated}. Rainfall, as one of the most critical meteorological parameters, significantly impacts transportation, agriculture, and daily social activities. Prolonged or intense rainfall can lead to devastating natural disasters such as landslides and floods. Consequently, timely and accurate rainfall monitoring is essential for safeguarding lives and property \cite{christofilakis2020rain, christofilakis2020earth, avanzato2020hydrogeological}.
	
	Traditional rainfall monitoring methods, including rain gauges, meteorological radars, and weather satellites, have notable limitations \cite{goldshtein2009rain, anagnostou1999uncertainty, lorenz2012hydrological}. Rain gauges and radars provide fine temporal resolution but are constrained by their installation density and susceptibility to errors like echo interference and scattering. Satellites, while offering broad coverage, lack the precision required for accurate rainfall intensity estimation due to indirect retrieval methods \cite{jurczyk2020quality, kidd2011global}. As a result, these conventional methods struggle to balance precision and spatial coverage. By contrast, ISAC networks, with their dense configurations of base stations (BS) and user nodes, offer the potential to improve both aspects, presenting significant advantages for rainfall monitoring.
	
	Research into leveraging wireless communication networks for rainfall classification has been steadily gaining momentum \cite{messer2006environmental, berne2007path, arnold1981rain, leijnse2007hydrometeorological, uijlenhoet2018opportunistic, itur2021propagation, ojo2008rain, ojo2015diurnal,kamga2019wireless}. The International Telecommunication Union's Radiocommunication Sector (ITU-R) has identified rain attenuation as a significant factor affecting frequencies above 5 GHz \cite{itur2021propagation}. This recognition has spurred a growing research focus on millimeter-wave bands, where rain-induced signal degradation becomes a critical concern. Additionally, the increasing availability of power measurement data from cellular operators has further enabled research into rainfall estimation using commercial microwave links, particularly those operating in the 10--30 GHz range \cite{messer2006environmental, berne2007path, leijnse2007hydrometeorological, uijlenhoet2018opportunistic}. Rainfall, as the predominant atmospheric factor causing microwave signal attenuation, primarily impacts signals through absorption and scattering. By establishing a robust understanding of the relationship between rainfall and signal attenuation, communication links can be repurposed as effective tools for rainfall monitoring. To facilitate this, power-law models are frequently employed to estimate path loss resulting from signal attenuation and to convert these losses into rainfall rates along the microwave link path \cite{atlas1977path, olsen1978arbrelation}. 

	While previous research has emphasized higher frequencies, the widespread adoption of sub-6 GHz bands in commercial 5G networks presents a new opportunity for rainfall detection. Leveraging these lower frequencies for environmental sensing could maximize the utility of existing 5G infrastructure, yielding both commercial and societal benefits. This realization has prompted growing interest in exploring the impact of rain attenuation within the sub-6 GHz spectrum. Recent findings suggest that rain attenuation in sub-6 GHz bands, while less pronounced, is far from negligible. For example, experiments conducted at 1.8 GHz demonstrated that mobile phones could reliably detect variations in received signal strength (RSS) caused by rain \cite{fang2016impact}. The attenuation patterns observed at these frequencies were found to align closely with those predicted for higher frequencies, underscoring the relevance of sub-6 GHz frequencies for rainfall estimation.
	
	Building on this, the first rainfall classification method using 4G/LTE RSS measurements achieved a classification accuracy of 96.7\% by extracting features such as instantaneous RSS, average RSS, and RSS variance, and applying a probabilistic neural network \cite{beritelli2018rainfall}. Further validation came from customized systems measuring S-band rain attenuation in Greece, which confirmed the significance of rain-induced signal attenuation at sub-6 GHz frequencies \cite{christofilakis2020rain}. Subsequent research demonstrated the accuracy of refined power-law relationships in characterizing sub-6 GHz rain attenuation \cite{sakkas2024harnessing,sakkas2024measuring}. Furthermore, recent work also explored the use of a poisson model to accurately characterize attenuation in 5 GHz microwave links \cite{ramos2024statistical}. These controlled experiments provided valuable insights into the characteristics of rain attenuation at lower frequencies, highlighting the potential for broader adoption of these methods.
	
	However, previous studies predominantly relied on RSS as the sole metric for rainfall analysis. Channel state information (CSI) offers a more detailed view of signal propagation. CSI has proven effective in applications like localization \cite{zhu2022intelligent,tong2021csi,zheng2019oparray},  channel prediction\cite{karami2013pilot}, human activity recognition \cite{wang2019joint}, and health monitoring \cite{soto2022survey}, suggesting its potential for capturing nuanced channel characteristics. To the best of our knowledge, CSI has not yet been applied for rainfall classification, presenting a novel opportunity to enhance sensing capabilities in sub-6 GHz frequencies.
	
\begin{figure}[t]
		\centering
		\includegraphics[width=3.5in]{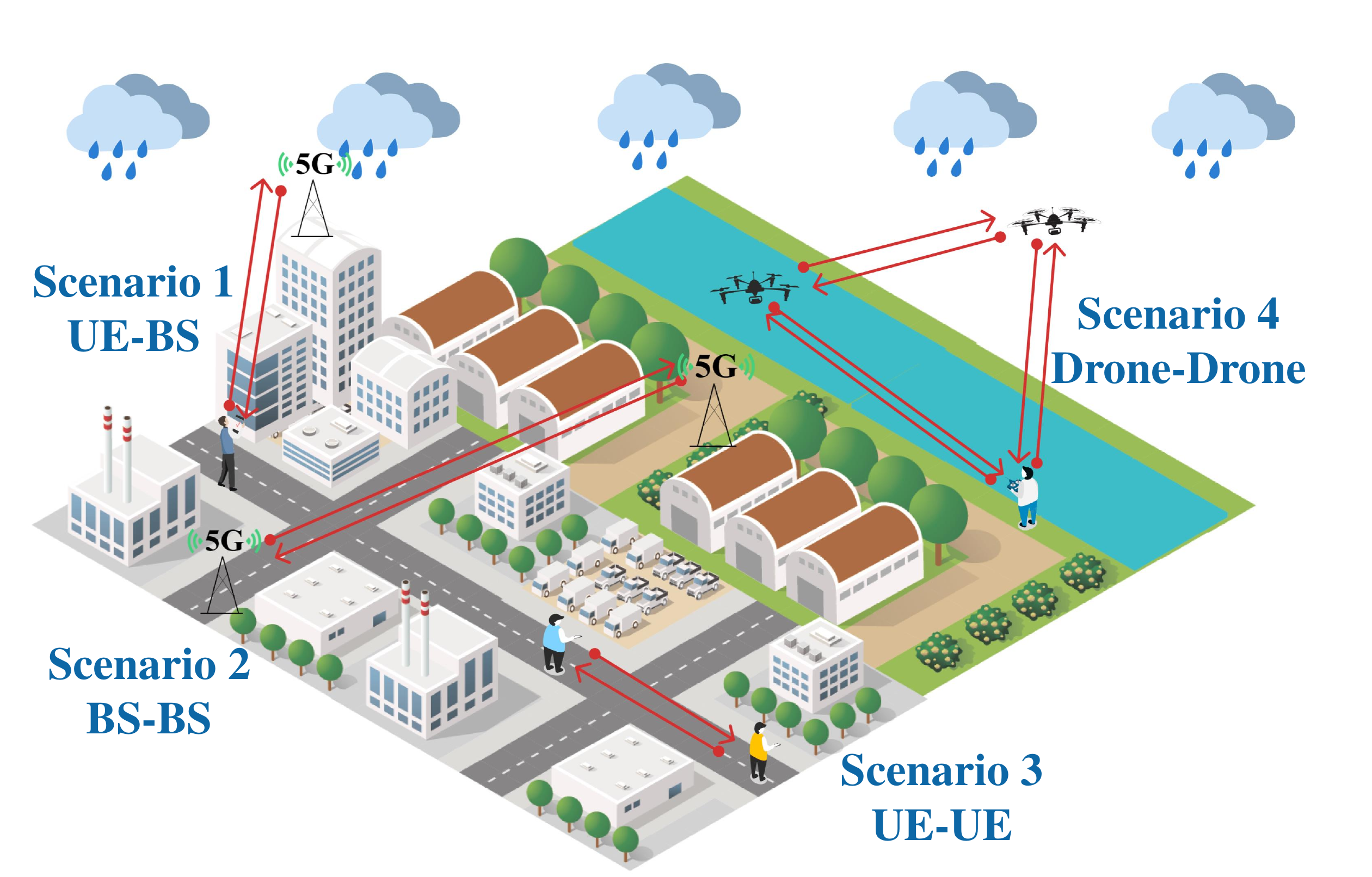}
		\caption{
			Application scenarios of the CSI-based smart rainfall gauge for urban and remote environments, demonstrating various deployment configurations for comprehensive rainfall monitoring:
			1) {\bf UE-BS:} Coverage of specific areas with large-scale data collection from multiple UEs.
			2) {\bf BS-BS:} Monitoring over extended regions through inter-base station communication.
			3) {\bf UE-UE:} Flexible and customizable configurations based on specific demands.
			4) {\bf Drone-Drone:} Mobile and capable of accessing hard-to-reach locations such as valleys or lakes.
		}			
		\label{fig:scenario}
	\end{figure}

	In this study, we leverage CSI to perform fine-grained measurements of communication channels under various rainfall conditions and develop a smart rainfall gauge based on CSI features, as illustrated in Fig.~\ref{fig:scenario}. Our system, utilizing commercially available devices, enables rainfall classification across multiple scenarios, including UE-to-BS, BS-to-BS, and even drone-based measurements for hard-to-reach areas like valleys or lakes. Unlike traditional rain gauges, which are fixed at specific locations and unsuitable for extreme environments, this approach facilitates rapid, scalable rainfall classification, particularly during severe weather events.
	
	Despite its potential, CSI-based rainfall classification faces challenges. Moisture accumulation on protective radomes during rainfall can form water films, introducing additional attenuation that alters antenna directivity and reflectivity, potentially causing overestimations of rainfall intensity \cite{moroder2020modeling, zhang2023precipitation, leijnse2008microwave}. Additionally, environmental factors such as temperature and humidity introduce variability in CSI measurements unrelated to rainfall, complicating consistent classification. To address these challenges, we incorporate temporal features from consecutive CSI snapshots, capturing nuanced variations that improve robustness. Our main contributions are as follows:
	
	\begin{itemize}
		\item \textit{Comprehensive sub-6 GHz Rain Attenuation Channel Measurement:} We established a system to measure RSS and CSI under varying rainfall conditions, analyzing parameters such as power delay profiles (PDP), multipath decay factors, and RMS delay spread to understand sub-6 GHz rain attenuation effects comprehensively.
		
		\item \textit{CSI-Based Rainfall Classification Framework:} We developed RainGaugeNet, a model leveraging multipath and temporal features from consecutive CSI snapshots, overcoming the limitations of single-time RSS-based measurements and enabling fine-grained rainfall classification.
		
		\item \textit{Robustness Validation Across Diverse Conditions:} Our dataset spans four environments and three rainfall intensities, demonstrating RainGaugeNet's classification accuracy of over 90\% in line-of-sight (LoS) scenarios and over 85\% in non-line-of-sight (NLoS) scenarios with only 20 seconds of CSI data.
		
	\end{itemize}
	
	The rest of this paper is organized as follows. Section~\ref{sec:SM} introduces the system model, describes the experimental setup and analysis of RSS and CSI under different rainfall conditions. Section~\ref{sec:CSI} presents the rainfall classification algorithm and experimental results. Finally, Section~\ref{sec:conclusion} concludes the study. Key variables are listed in Table~\ref{NOTATIONS}.
	
	\begin{table*}[h]
		\centering
		\caption{Notations of Important Variables}\label{NOTATIONS}
		\begin{tabular}{@{}l l @{\hspace{1cm}} l l@{}}
			\toprule
			\textbf{Notation} & \textbf{Definition} & \textbf{Notation} & \textbf{Definition} \\
			\midrule
			$\Delta f$ & Subcarrier spacing & $N_\mathrm{p}$ & Number of downlink propagation paths \\
			\midrule
			$\delta_{t,p}$ & Time of arrival of the $p$-th path at time $t$ & $\alpha_{t,m,p}$ & Complex gain of the $p$-th path on $m$-th antenna \\
			\midrule
			$X$ & Known pilot symbols (assumed $X = 1$) & $H_{t,m,n}$ & CFR of the $n$-th subcarrier on $m$-th antenna \\
			\midrule
			${W}_{t,m,n}$ & Unresolvable multipath interference and noise & $\widehat{h}_{t,n}$ & Estimated average channel impulse response \\
			\midrule
			$T_\mathrm{s}$ & Sampling interval & $\tau_n$ & Time delay of $n$-th tap, $\tau_n = n T_\mathrm{s}$ \\
			\midrule
			$P_n$ & Power of the $n$-th tap in STDL model & $N_\mathrm{T}$ & Total number of taps (length of observation window) \\
			\midrule
			$N_\mathrm{L}$ & Total number of identified multipath components & $X_\mathrm{PDP}$ & Random variable in power-law decay model \\
			\midrule
			$n_\mathrm{PDP}$ & Decay factor in power-law decay model & $P_{n}^{\text{meas}}$ & Measured power data at the $n$-th bin \\
			\midrule
			$P_{n}^{\text{calc}}$ & Calculated power data at the $n$-th bin & $\sigma_\mathrm{RMSE}$ & RMSE between measured and calculated data \\
			\midrule
			$P_{\mathrm{th}}$ & Detection threshold for multipath components & $P_{\mathrm{max}}$ & Peak power of the PDP \\
			\midrule
			$N_0$ & Noise floor & $\gamma_\mathrm{P}$ & Relative power threshold relative to $P_\mathrm{max}$ \\
			\midrule
			$\gamma_\mathrm{N}$ & Power threshold relative to noise floor & $P_r$ & Total received multipath power \\
			\midrule
			$k_l$ & the tap index of $l$-th path among multipath components & $\tau_\mathrm{RMS}$ & RMS delay spread \\
			\midrule
			$\bar{\tau}$ & Mean delay of multipath components &  &  \\
			\bottomrule
		\end{tabular}
	\end{table*}

	\section{Sub-6 GHz Rainfall Channel Measurement and Analysis}
	\label{sec:SM}
	In this section, we introduce the system model of the sub-6 GHz rainfall channel, followed by a presentation of the sub-6 GHz rainfall measurement system. We then analyze the RSS and CSI under varying rainfall intensities.
	\subsection{System Model}	
	We consider a downlink communication scenario in a cellular orthogonal frequency division multiplexing (OFDM) system, where a connection is established between a UE and a BS. The BS transmits using beamforming to direct signals to the UE, employing an antenna array. We assume that the downlink frequency band has a subcarrier spacing of $\Delta f$ and that the antenna spacing is half of the wavelength. At time instant $t$, the channel frequency response (CFR) at $n$-th subcarrier, on the $m$-th antenna of the UE-BS link can be expresed as
	\begin{equation}
		H_{t,m,n} = \sum_{p=1}^{N_\mathrm{p}} \alpha_{t,m,p} e^{-j 2 \pi  n \Delta f \delta_{t,p}} 
	\end{equation}		
	where $N_\mathrm{p}$ denotes the number of downlink channel propagation paths, and those $\alpha_{t,m,p}$ and $\delta_{t,p}$ represent the complex gain and time of arrival of the $p$-th path, respectively. In this formulation, the array response at the $m$-th antenna is incorporated into the complex gain. Consequently, the received signal of UE at $n$-th subcarrier, on the $m$-th antenna is expressed as 
	\begin{equation}
		Y_{t,m,n} = {H}_{t,m,n}{X} + {W}_{t,m,n},
		\label{eq:ytf}
	\end{equation}
	where $X$ denotes the known pilot symbols, and for simplicity, we assume ${X = 1}$. The term ${W}_{t,m,n}$ encompasses the other unresolvable multipath interference and additive white Gaussian noise.  
	
	\subsection{Sub-6 GHz Rainfall Channel Measurement System}
	\begin{figure}[h]
		\centering
		\includegraphics[width=3.5in]{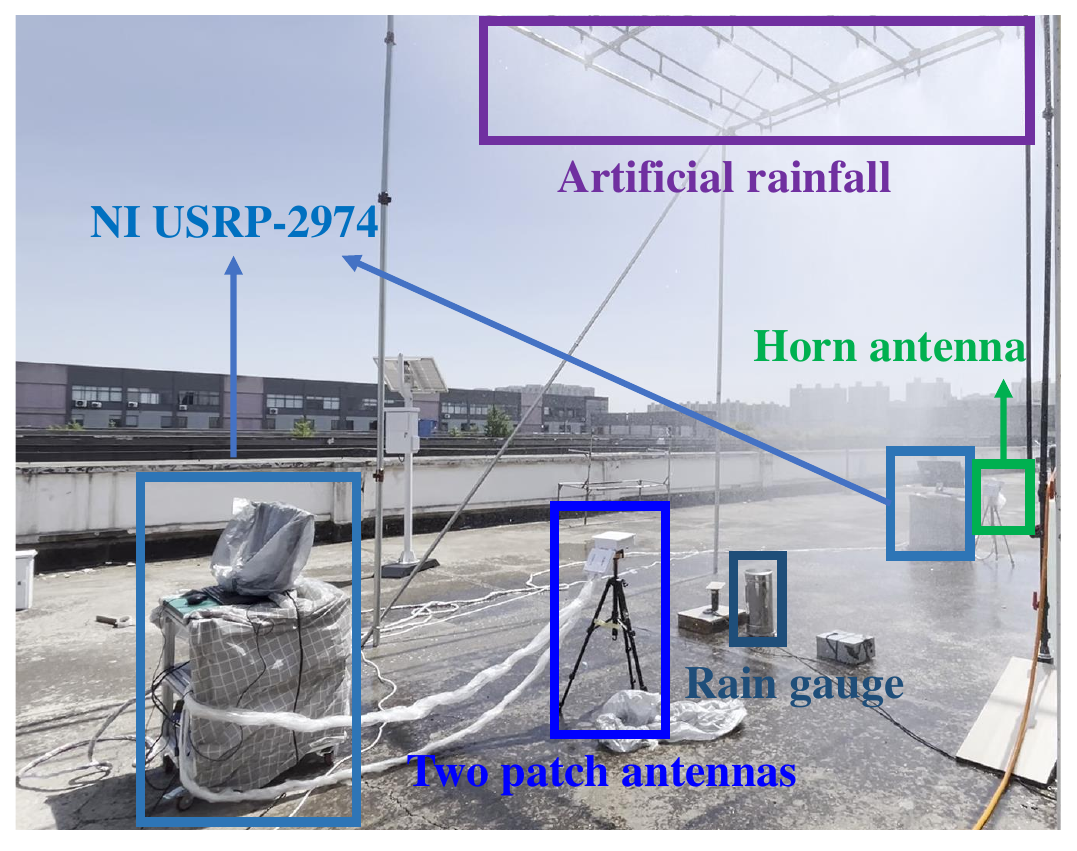}
		\caption{Experimental field layout of the sub-6 GHz rainfall measurement system, with the hardware modules connected as shown in Fig.~\ref{fig:system}.}
		\label{fig:real}
	\end{figure}
	
	\begin{figure*}[h]
		\centering
		\includegraphics[width=7in]{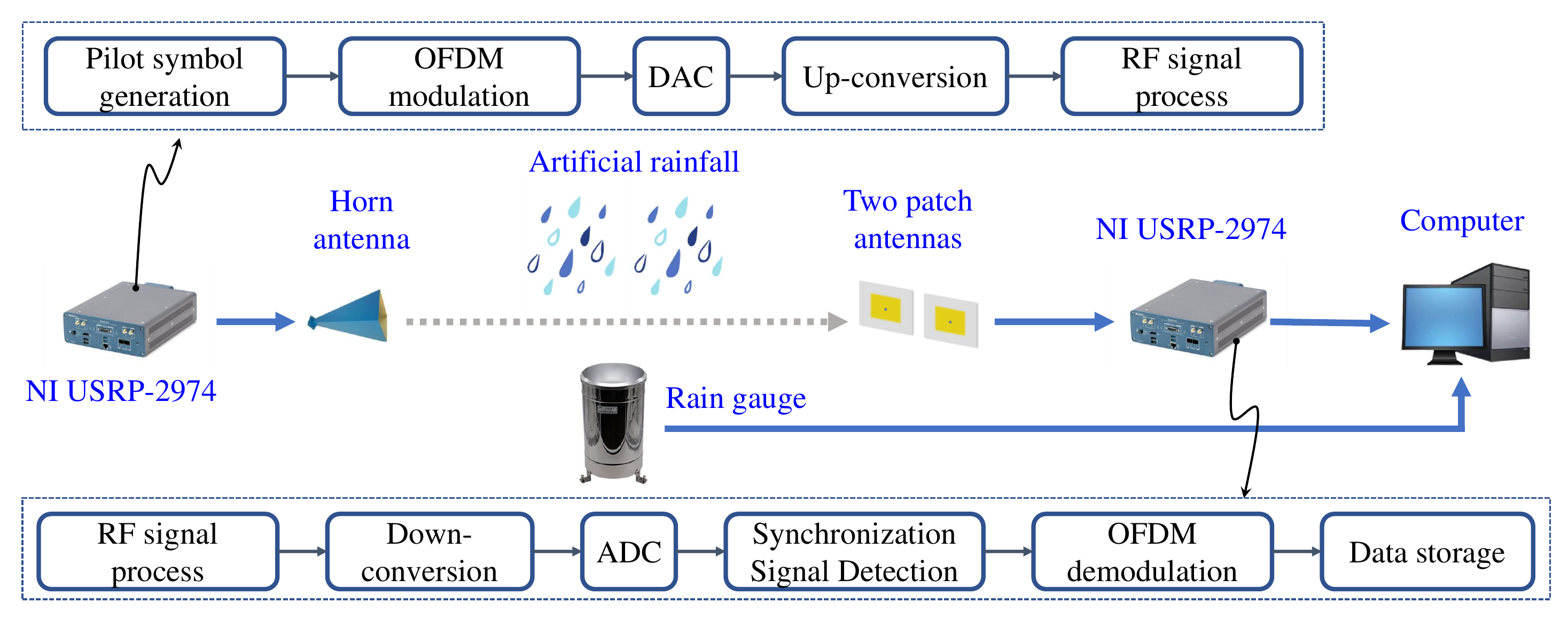}
		\caption{Overview of the sub-6 GHz rainfall measurement system, illustrating the main hardware components and the internal data stream of the NI USRP-2974.}
		\label{fig:system}
	\end{figure*}
	The sub-6 GHz rainfall measurement system, established in Nanjing, China, is illustrated in Fig.~\ref{fig:real}, while the detailed experimental setup is shown in Fig.~\ref{fig:system}. The system includes an NI USRP-2974 transmitter equipped with a horn antenna, operating at 2.8 GHz with 100 MHz OFDM modulation. The data stream undergoes up-conversion and other necessary adjustments before being transmitted through the antenna. One subframe, containing two time slots, is captured approximately every second.        
	
	The receiver, designed to emulate a smartphone, is also based on an NI USRP-2974 and equipped with two patch antennas. It handles synchronization signal detection and other related tasks. A common clock shared by the transmitter and receiver is linked via an SMA cable to ensure synchronization. Both the transmitter and receiver are positioned at a height of 1.5 meters (m), with a separation distance of 7 m and a zero-degree elevation angle.
	
	The rainfall simulation system generates spatially and temporally uniform rainfall at two intensities---5 millimeters (mm) and 20 mm---within a 5-minute period. A rain gauge is used to measure the rainfall amount to verify the accuracy of the artificial rainfall and to prevent errors during the process. After each rainfall event, the measurements are recorded, and the system is reset to ensure consistent conditions.
	
	Once the artificial rainfall concludes, the collected data, including rainfall measurements and received signals, are transferred to a computer for offline analysis. In the following subsections, we analyze the RSS and CSI under varying rainfall conditions in the LoS scenario, maintaining wind speeds below 5 m/s to minimize wind-induced effects.

	\subsection{RSS Model and Characteristics}
	Rain attenuation estimation for radio signals can be broadly categorized into two approaches: physical methods and empirical methods.
	
	Physical methods aim to reconstruct theoretical models of the attenuation process, delving deeply into the physical mechanisms involved in the propagation of radio waves through the atmosphere. These methods often require complex numerical analyses and intricate mathematical formulations, making them computationally intensive. Furthermore, the required input parameters for these theoretical models are not always readily available, which limits their practical applicability \cite{kestwal2014prediction}.
	
	In contrast, empirical models are based on measurement databases obtained from various climatic zones. Their primary advantage lies in their relatively simple mathematical formulations, making them convenient for practical application. Consequently, empirical methods are the most widely adopted approach for estimating rain attenuation \cite{kestwal2014prediction, samad2021survey, mello2012unified, garcialopez1981modified}. 
	Among these, the method recommended by the ITU-R \cite{itur2021propagation} is the most widely used, where \cite[Eq. (32)--(33)]{itur2021propagation} are used to obtain the specific rain attenuation.
	
\begin{figure} 
		\centering
		\includegraphics[width=0.5\textwidth,trim=0cm 0cm 0cm 0cm,
		clip]{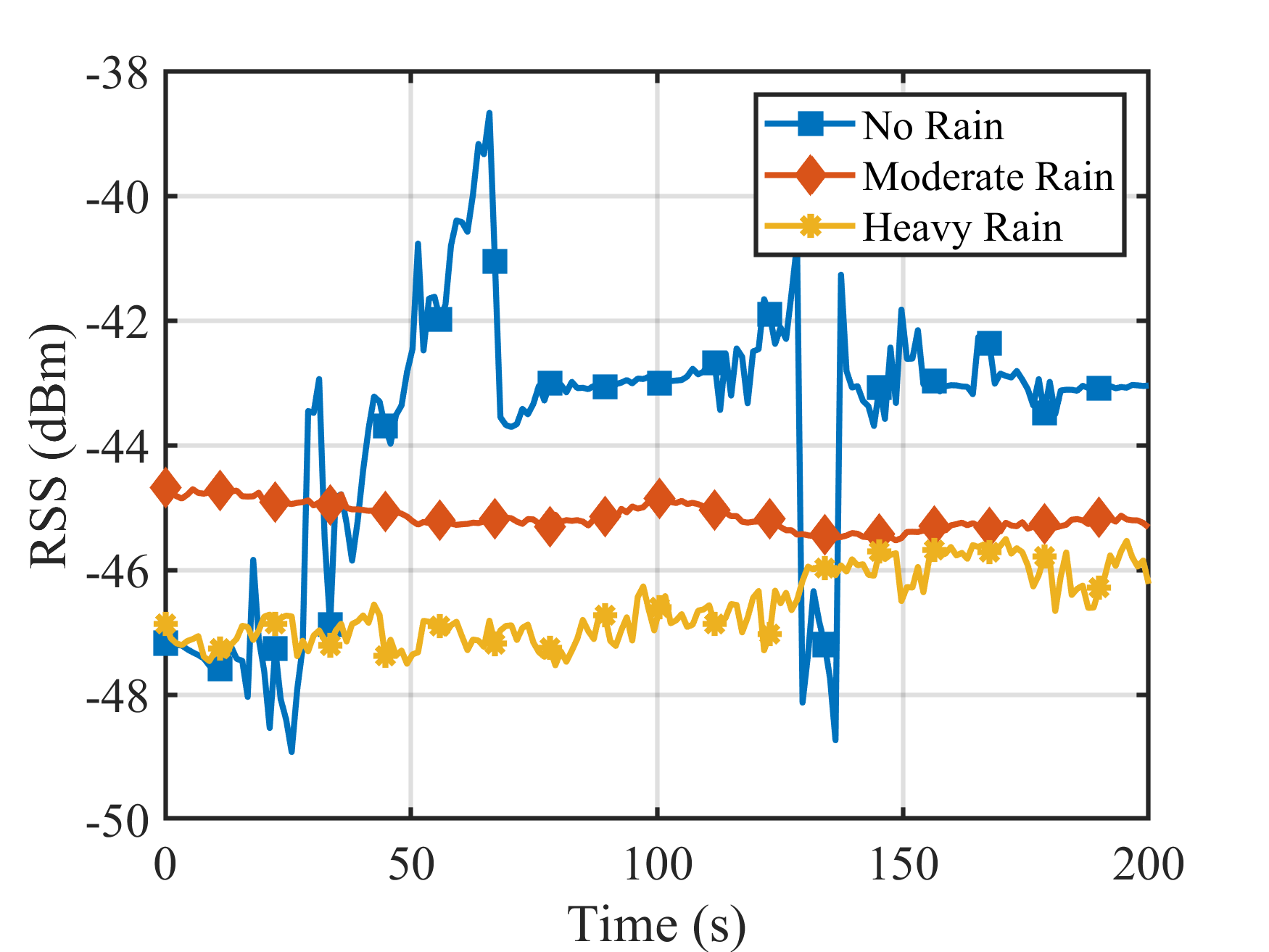}
		\caption{Instantaneous RSS values under different rainfall intensities.}
		\label{fig:shunshi}
	\end{figure}

	RSS is often used to distinguish different rainfall intensities. For each received subframe of data, we compute the RSS from all its IQ data. The instantaneous RSS values for three different rainfall scenarios are shown in Fig.~\ref{fig:shunshi}. It is evident that rain attenuation affects sub-6 GHz signals, with observable differences in RSS between varying rainfall intensities. Specifically, using \cite[Eq. (32)--(33)]{itur2021propagation}, the calculated rain attenuation for moderate and heavy rain is 0.0047 dB and 1.33 dB, respectively. However, the actual average attenuations are 1.8647 dB and 3.2814 dB, respectively. Since the ITU-R model is primarily derived for high frequencies, applying it to sub-6 GHz introduces deviations. Consequently, differences in actual rain attenuation across rainfall conditions are neither substantial nor consistent, complicating the task of distinguishing between rainfall intensities.
	
\begin{figure} 
		\centering
		\includegraphics[width=0.5\textwidth,trim=0cm 0cm 0cm 0cm,
		clip]{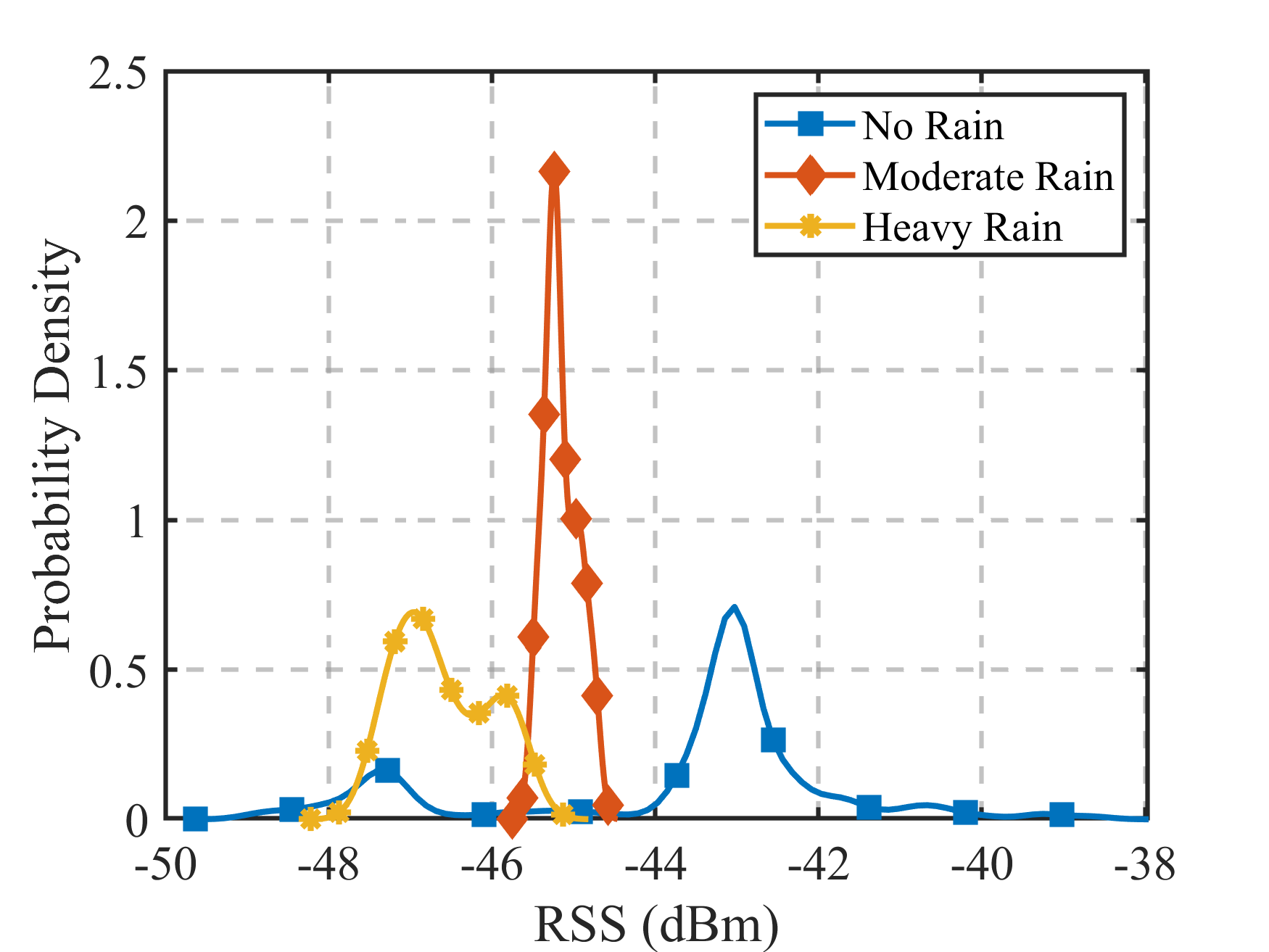}
		\caption{PDF of instantaneous RSS under different rainfall intensities.}
		\label{fig:pdf}
	\end{figure}

\begin{figure} 
		\centering
		\includegraphics[width=0.5\textwidth,trim=0cm 0cm 0cm 0cm,
		clip]{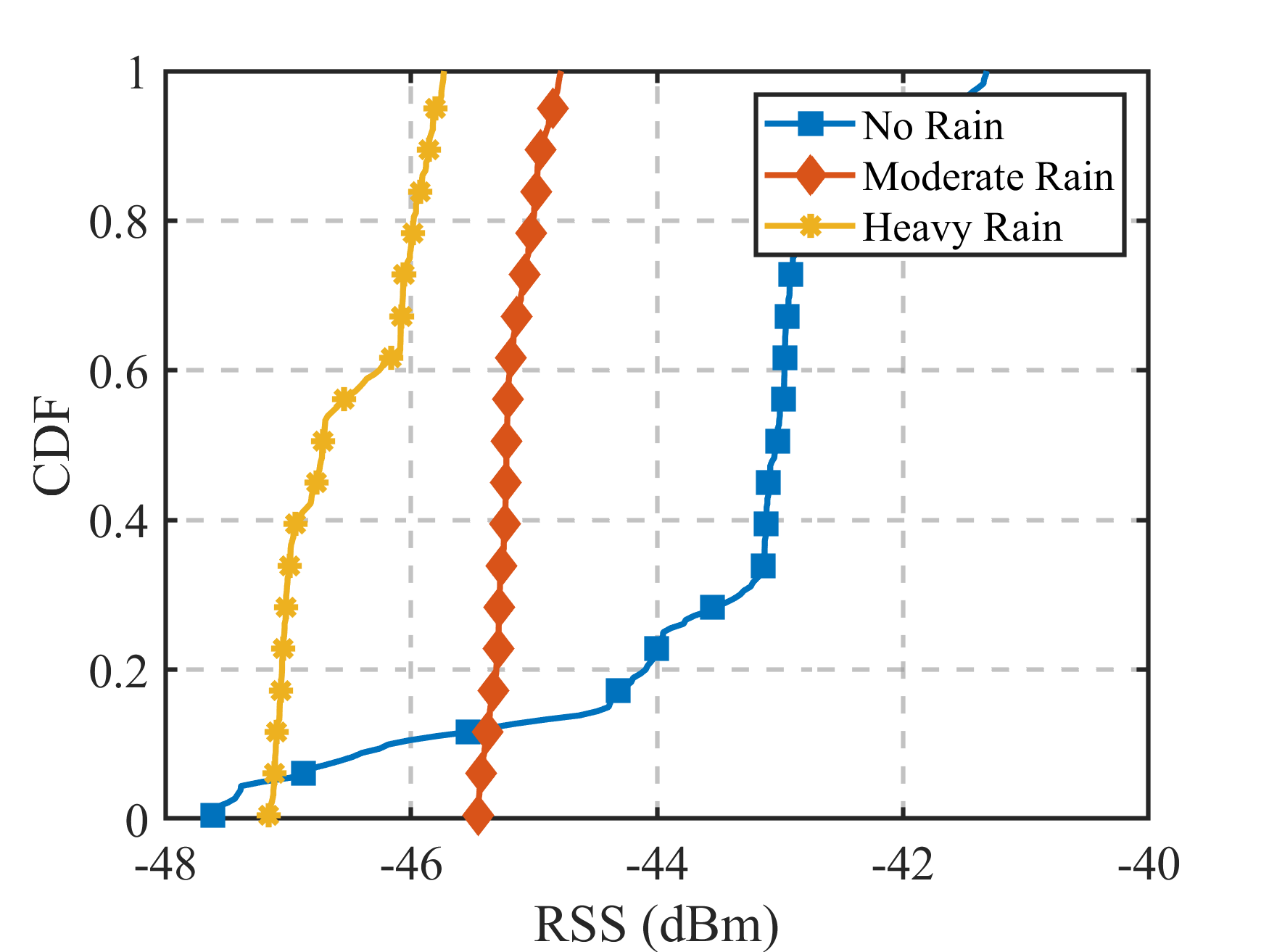}
		\caption{CDF of sliding window mean RSS (window size: 20) under different rainfall intensities.}
		\label{fig:meancdf}
	\end{figure}

\begin{figure} 
		\centering
		\includegraphics[width=0.5\textwidth,trim=0cm 0cm 0cm 0cm,
		clip]{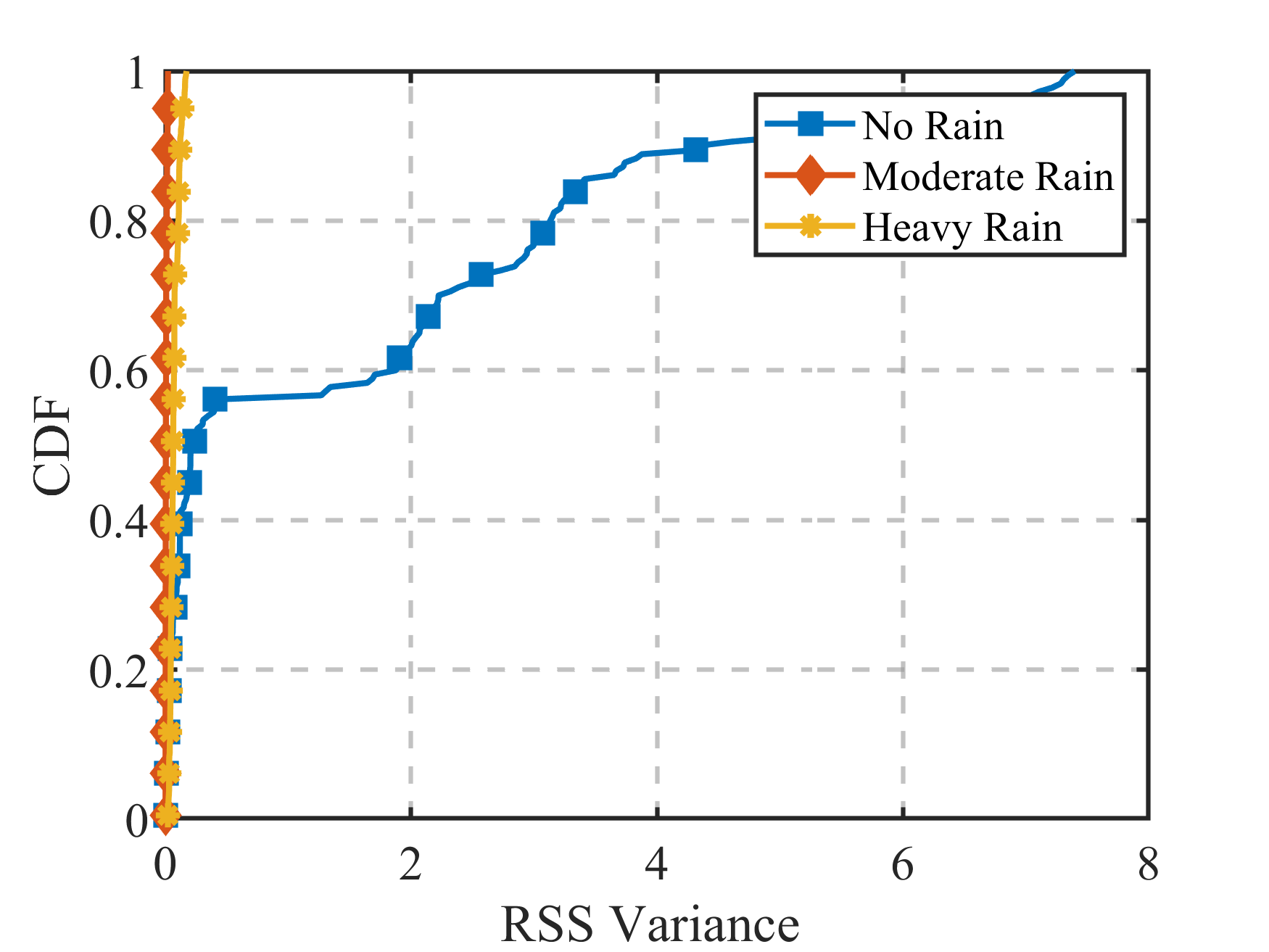}
		\caption{CDF of sliding window RSS variance (window size: 20) under different rainfall intensities.}
		\label{fig:varcdf}
	\end{figure}

	Inspired by the classification of rain attenuation in LTE/4G mobile signals, we present the probability density function (PDF) distributions of RSS under different rainfall conditions in Fig.~\ref{fig:pdf}. The mean and variance of RSS vary across different rainfall intensities. To compute these statistics, a sliding window of size 20 is applied, covering approximately 20 seconds of rainfall data. The cumulative distribution function (CDF) of the RSS mean and variance are shown in Fig.~\ref{fig:meancdf} and Fig.~\ref{fig:varcdf}, respectively.
	
	Analysis of RSS measurements under different rainfall conditions reveals the following:
	\begin{itemize}
		\item \textbf{O1:} The CDF of the RSS mean indicates that no-rain conditions generally yield higher RSS values compared to rain conditions. However, the RSS signal under no-rain conditions is unstable, complicating the task of accurately determining precise rainfall intensity. Furthermore, the RSS means for moderate and heavy rainfall are quite similar, presenting additional classification challenges and increasing susceptibility to interference in practical applications.
		
		\item \textbf{O2:} The variance of RSS decreases as rainfall intensity increases, aligning with previous work \cite{beritelli2018rainfall}. This observation is attributed to the impact of two types of multipath components influenced by rainfall, which depend on the surrounding environment and rainfall intensity:
		\begin{itemize}
			\item {\bf Type 1 Multipath:} Originates from scatterers such as raindrops along the original LoS path. Given the relatively short distance between the transmitter and receiver, only multipath components close to the LoS path (undergoing few reflections) retain significant energy. In contrast, paths with multiple reflections suffer multiplicative fading and rain attenuation, reducing their energy. Despite the random distribution of raindrops, the large number of drops between the transmitter and receiver results in statistically stable signal energy, leading to low RSS variance. However, these multipath components can significantly affect the delay spread, a phenomenon validated in the next subsection.
			
			\item {\bf Type 2 Multipath:} Originates from scatterers in the surrounding environment. Variability introduced by moving objects impacts RSS. Due to long propagation paths and energy losses from reflections, these multipath components typically exhibit lower energy than Type 1. Consequently, they are easily overwhelmed by even slight rain attenuation, reducing variance as rainfall intensity increases.
		\end{itemize}
		
	\end{itemize}
	
	These observations highlight the inherent variability and challenges of using RSS measurements for rainfall classification. Both the RSS mean and variance lack consistent indicators to reliably differentiate between rainfall conditions. This inconsistency underscores the need for a more granular metric. Consequently, we turn to CSI, which provides fine-grained rainfall characteristics, enabling accurate and reliable feature extraction.
	
	\subsection{PDP Model and Characteristics}
	
	In this subsection, we investigate multipath propagation by analyzing the PDP derived from the CSI. We use the least squares algorithm for channel estimation and compute the average estimated CFR across all antennas to enhance the robustness of the estimation process. Applying the inverse discrete fourier transform, we obtain $\widehat{h}_{t,n}$. Subsequently, the PDP is calculated as
	\begin{equation}
		\text{PDP}_{t,n} = |\widehat{h}_{t,n}|^2.
		\label{eq:PDP}
	\end{equation}
	
	The traditional model for characterizing small-scale multipath propagation channels is the stochastic tapped delay line model \cite{hashemi1993impulse,cassioli2002ultra,lee2010uwb}, where the delay axis is divided into small time bins. Each bin may or may not contain a multipath component, with the size of each bin corresponding to the resolution of the measurement system. In our study, the resolution of the system is defined by the sampling interval, denoted as $T_\mathrm{s}$, which is 10 nanoseconds (ns). In this model, the received power is represented as discrete tap coefficients at specific delay values. Specifically, at each discrete tap within the total number of taps, or equivalently, the length of the observation window for the PDP, denoted as $N_\mathrm{T}$, the corresponding power at time delay $\tau_n = n T_\mathrm{s}$ is represented by $P_n$.

	For our analysis, we selected $N_\mathrm{T} = 40$, corresponding to a delay span of 400 ns, which is sufficient to capture the key multipath components. Typically, PDP decay follows either an exponential \cite{hashemi1993impulse} or power-law \cite{cassioli2002ultra} distribution. In LoS scenarios with strong reflections, the power-law decay model has been found to better fit the measurement data \cite{karedal2007measurement, ai2015power,ren2024time}.
	The power-law decay is described as
	\begin{equation}
		P_n = \frac{a}{\tau_n^{n_{\mathrm{PDP}}}},
	\end{equation}  
	where $a$ is a constant, and $n_{\mathrm{PDP}}$ is the decay factor. Taking the logarithmic form and accounting for the overall uncertainty, the power-law decay model is transformed into
	\begin{equation}
		10 \log_{10} P_n = \eta_0 - n_{\mathrm{PDP}} \cdot 10 \log_{10} \tau_n + X_{\mathrm{PDP}},
		\label{eq:powerlow}
	\end{equation}
	where $10 \log_{10} P_n$ is linearly related to $10 \log_{10} \tau_n$, $\eta_0 = 10 \log_{10} a$ is a constant, and $X_\mathrm{PDP} \sim {\cal N}(0, \sigma^2)$ is a random variable normally distributed. The parameter $\sigma$ represents the overall uncertainty, which may arise from measurement inaccuracies, model fitting errors, or other sources of variability in the data.
	
	To analyze attenuation behavior, all taps of each PDP are superimposed after power and delay normalization relative to the first tap. The PDP data is fitted using \eqref{eq:powerlow}. The root mean squared error (RMSE) between measured data $P_{n}^{\text{meas}}$ and the calculated data $P_{n}^{\text{calc}}$ is computed as
	\begin{equation} 
		\sigma_\mathrm{RMSE} = \sqrt{\frac{1}{N_\mathrm{T}} \sum_{n=1}^{N_\mathrm{T}} \left(P_{n}^{\text{meas}} - P_{n}^{\text{calc}}\right)^2}.
	\end{equation}

	We use the least mean square error method to fit average PDP curves for the first 400 ns of data. For each rainfall intensity, 3,000 randomly selected data sets were used for fitting. The final fitting results, demonstrating excellent accuracy due to low RMSE values, are presented in Table~\ref{tab:pdp_fitting}.

	\begin{itemize}
		\item[\textbf{O3:}] The attenuation channel conforms to a power-law decay model, with the decay factor $n_{\mathrm{PDP}}$ decreasing as rainfall intensity increases. This suggests that attenuation in LoS paths occurs more rapidly than in NLoS paths under increased rainfall. This phenomenon may arise because some NLoS paths partially escape rain exposure.
		
	\end{itemize}

	\begin{table} 
		\centering
		\normalsize
		\caption{Fitting Results for PDP under Different Rainfall Conditions}
		\label{tab:pdp_fitting}
		\begin{tabular}{cccc}
			\toprule
			\textbf{Parameter} & \textbf{No Rain} & \textbf{Moderate Rain} & \textbf{Heavy Rain} \\
			\midrule
			$\eta_0$       & 0.6   & 0.94  & 0.64  \\
			$n_\mathrm{PDP}$       & 1.52    & 1.49  & 1.41  \\
			$\sigma_\mathrm{RMSE}$ & 4.99    & 3.50  & 1.60  \\
			\bottomrule
		\end{tabular}
	\end{table}

	\subsection{Multipath Parameter Extraction and Characteristics}

	\begin{figure}
		\centering
		\includegraphics[width=0.5\textwidth, trim=0cm 0cm 0cm 0cm, clip]{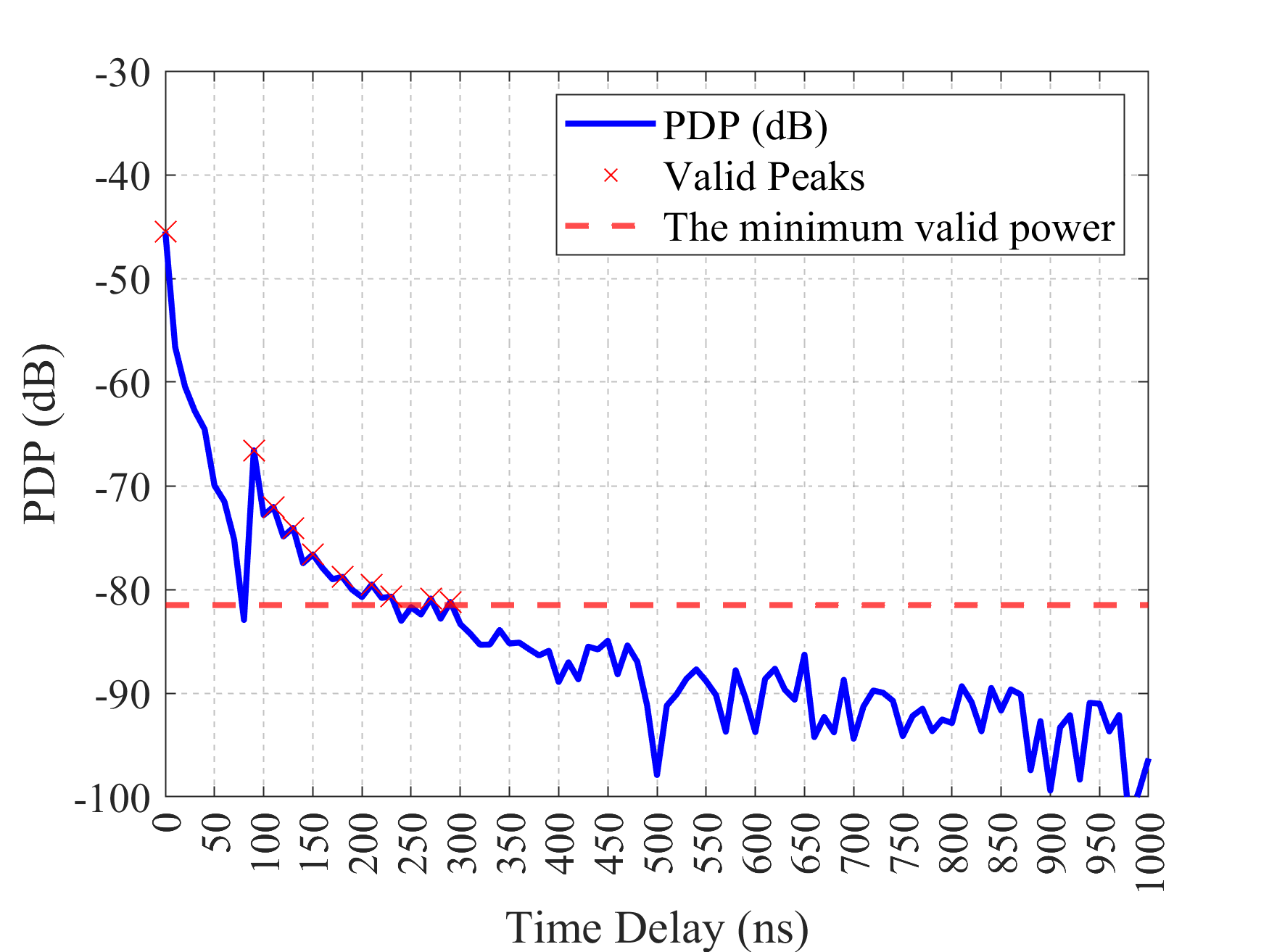}
		\caption{
			An example of PDP, where the multipath identification algorithm is applied to the measured PDP, resulting in the determination of a detection threshold with red dashed line. The valid MPCs that exceed this threshold are marked with red crosses.
		}
		\label{fig:PDP}
	\end{figure}
	
	In this subsection, we analyze multipath propagation by extracting parameters from the PDP. The PDP serves as a fundamental tool for identifying multipath components (MPCs) and analyzing key channel characteristics, such as delay dispersion. A robust multipath identification algorithm was implemented to extract significant multipath features while mitigating noise interference \cite{ren2024time,sang2024measurement}.
	
	Significant MPCs are identified based on a predefined power threshold $P_{\mathrm{th}}$, ensuring the reliability of the detected components based on the noise floor and system parameters \cite{ren2024time,sang2024measurement}. The detection threshold is computed as
	\begin{equation}
		P_{\mathrm{th}} = \max\left(P_{\mathrm{max}} - \gamma_\mathrm{P} , N_0 + \gamma_\mathrm{N}\right),
		\label{eq:threshold}
	\end{equation}
	where $P_{\mathrm{max}}$ represents the peak power of the PDP, $\gamma_\mathrm{P}$ is the relative power threshold (set to 40 dB in our study), $N_\mathrm{0}$ is the noise floor, and $\gamma_\mathrm{N}$ is the threshold relative to $N_\mathrm{0}$ (set to 10 dB). The total received power from the identified multipath components is calculated as
	\begin{equation}
		P_r = \sum_{l=1}^{N_\mathrm{L}} P_{k_l},
		\label{eq:total}
	\end{equation}
	where $k_l$ is the tap index of the $l$-th identified path and $N_\mathrm{L}$ represents the total number of detected MPCs. The RMS delay spread characterizes the temporal dispersion of MPCs in wireless channels, offering insights into signal dispersion in multipath environments. It is defined as
	\begin{equation}
		\tau_{\mathrm{RMS}} = \sqrt{\frac{\sum_{l=1}^{N_\mathrm{L}} (\tau_{k_l} - \bar{\tau})^2 P_{k_l}}{\sum_{l=1}^{N_\mathrm{L}} P_{k_l}}},
		\label{eq:RMS}
	\end{equation}
	where $\bar{\tau}$ is the mean delay
	\begin{equation}
		\bar{\tau} = \frac{\sum_{l=1}^{N_\mathrm{L}} \tau_{k_l} P_{k_l}}{\sum_{l=1}^{N_\mathrm{L}} P_{k_l}}.
	\end{equation}
	
	\begin{figure}[!t]
		\centering
		\includegraphics[width=0.5\textwidth, trim=0cm 0cm 0cm 0cm, clip]{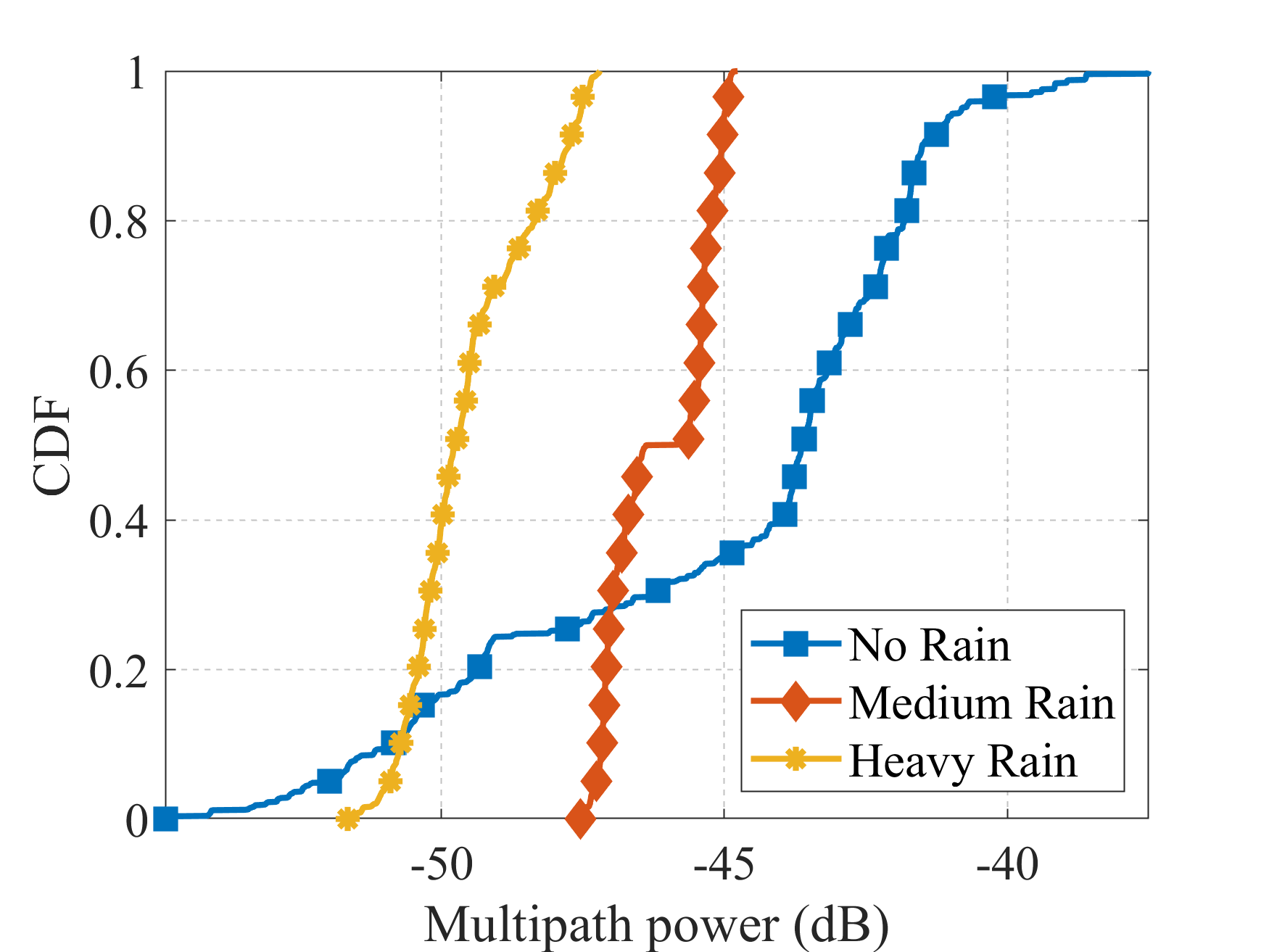}
		\caption{
			The CDF of the total received multipath power, calculated as the sum of the energies of all detected MPCs using the algorithm in \cite{ren2024time} and computed by \eqref{eq:total}.
		}
		\label{fig:power}
	\end{figure}
	
	\begin{figure}[!t]
		\centering
		\includegraphics[width=0.5\textwidth, trim=0cm 0cm 0cm 0cm, clip]{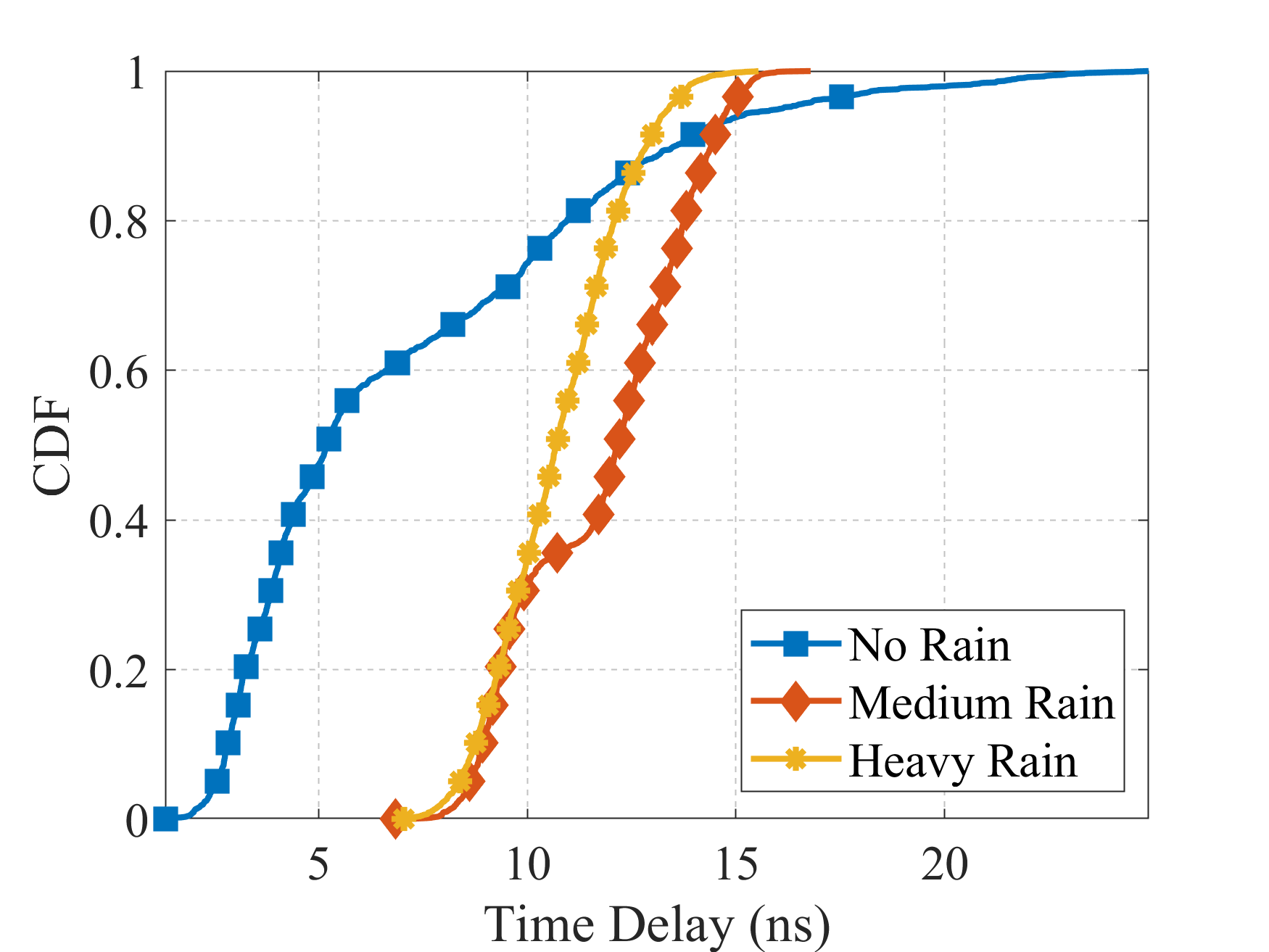}
		\caption{
			The CDF of the RMS delay spread, derived from the detected MPCs using the algorithm in \cite{ren2024time} and computed by \eqref{eq:RMS}.
		}
		\label{fig:RMS}
	\end{figure}
	
	\begin{table*}
		\centering
		\normalsize
		\caption{Multipath Parameters of Different Rainfall Intensities}
		\label{tab:pdpmutipath}
		\begin{tabular}{lccc}
			\toprule
			\textbf{Parameter}             & \textbf{No Rain} & \textbf{Moderate Rain} & \textbf{Heavy Rain} \\
			\midrule
			Multipath Power (dB)         & -43.91  & -46.03  & -49.50  \\
			Maximum Path Power (dB)      & -43.96  & -46.12  & -49.58  \\
			RMS Delay Spread (ns)          & 6.41  & 11.77 & 10.70 \\
			\bottomrule
		\end{tabular}
	\end{table*}
	
	The multipath parameters were extracted from the signals for three rainfall intensities, and the relevant parameters were calculated. To ensure robustness, the top and bottom 10\% of the data were excluded, and the mean values were computed from the remaining data, as shown in Table~\ref{tab:pdpmutipath}. The extraction of multipath parameters significantly mitigates the impact of noise, as evidenced by comparing rain attenuation values derived from multipath power to those obtained directly from IQ data. Specifically, rain attenuation values for moderate and heavy rain, measured at 2.12 dB and 5.59 dB, respectively, are markedly higher when derived from IQ data. This distinction highlights the utility of multipath parameter extraction in improving the accuracy and reliability of rainfall classification. However, as illustrated in Fig.~\ref{fig:power}, establishing a definitive threshold for rainfall classification remains challenging due to the large variability in energy during no-rain conditions.
	
	RMS delay spread provides crucial insights into the multipath characteristics of the channel. The CDF of the RMS delay spread is shown in Fig.~\ref{fig:RMS}. As outlined earlier, the two types of multipath components introduced by rainfall affect the delay spread differently. According to Table~\ref{tab:pdpmutipath}, a comparison of the maximum path power and multipath power reveals that the majority of the signal energy is concentrated in the LoS path, while the NLoS paths are relatively weaker. This observation indicates that Type 1 Multipath, which directly arises from scattering by raindrops near the LoS path, tends to experience fewer reflections and typically possesses higher energy compared to other multipath components. According to \eqref{eq:RMS}, this higher energy results in a greater weight in the RMS delay spread calculation, implying that the RMS delay spread is primarily determined by Type 1 Multipath. Therefore, the RMS delay spread under moderate rain is larger than that observed under no rain.
	
	As rainfall intensity increases, the number of raindrops acting as scatterers also increases, confining Type 1 Multipath components to paths closer to the LoS path—otherwise, they would not reach the receiving antenna. Consequently, the scattering effects of Type 1 Multipath are reduced, bringing the RMS delay spread down to 10.7 ns under heavy rain. However, the relationship between the RMS delay spread for  moderate and heavy rain remains uncertain due to the randomness of signal reflections from raindrops. Key observations from our study include:
	
	\begin{itemize}
		\item[\textbf{O4:}] As rainfall increases, there is significant attenuation of multipath energy. Identifying multipath components helps to mitigate noise, improving the detectability of rainfall attenuation compared to using IQ data alone. Additionally, LoS paths scattered by rainfall generate NLoS paths, leading to an increase in RMS delay spread. Interestingly, under heavy rainfall, NLoS paths experience less attenuation, resulting in a slight decrease in RMS delay spread compared to moderate rainfall conditions.
	\end{itemize}
	
	\section{CSI-Based Intelligent Rain Gauge}
	\label{sec:CSI}
	The analysis above demonstrates that multipath environment indicators, such as multipath power and RMS delay spread, exhibit notable variations under different rainfall intensities. However, relying on a single metric for precise classification of rainfall intensity remains challenging. In contrast, the CSI matrix contains a wealth of distinguishable features that reflect the characteristics of rainfall at varying intensities. This suggests that while individual metrics may lack sufficient accuracy, the CSI matrix as a whole holds valuable information for rainfall classification.
	
	Based on this insight, we propose a CSI-based intelligent rain gauge. This system leverages a rainfall classification network designed to extract features from the CSI matrix under diverse rainfall conditions. The network is specifically trained to analyze and interpret these features to determine rainfall intensity, requiring only 20 seconds of CSI measurement data for accurate classification.

	\subsection{Network Architecture}

	\begin{figure*}[h]
		\centering
		\includegraphics[width=\textwidth,trim=0cm 0cm 0cm 0cm, clip]{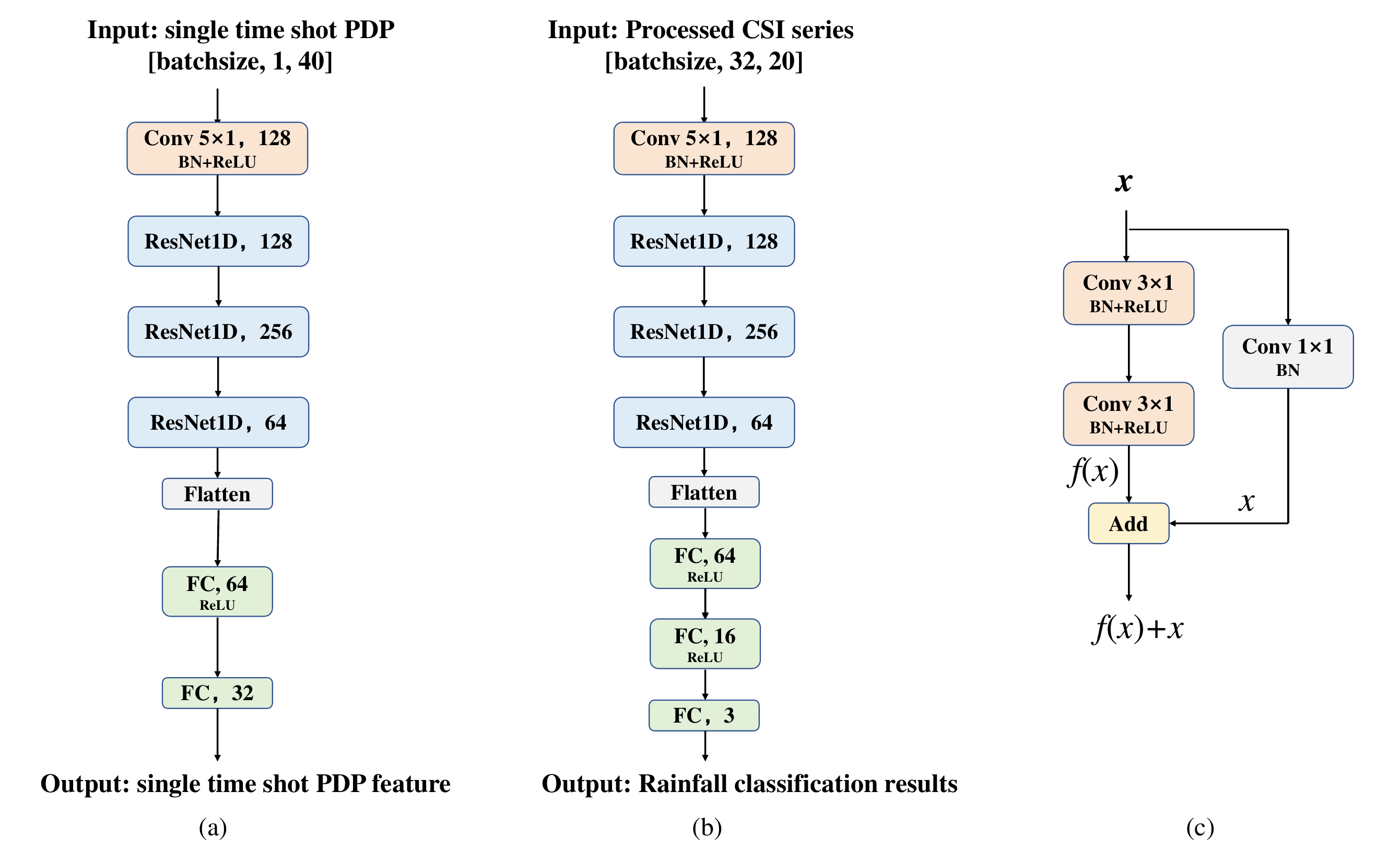}
		\caption{Overview of the RainGaugeNet algorithm architecture:
			(a) structure for extracting spatial features from individual Power Delay Profiles (PDPs) within RainGaugeNet,
			(b) structure for extracting temporal features from data processed over continuous 20 seconds by (a),
			(c) configuration of the ResNet1D component used within RainGaugeNet.}
		\label{fig:net}
	\end{figure*}

	Considering that rainfall impacts both the spatial and temporal characteristics of multipath signals, we developed \textbf{RainGaugeNet}, a robust neural network architecture tailored for classifying rainfall intensity (Fig.~\ref{fig:net}). To capture spatiotemporal features effectively, data preprocessing converts CSI into PDP, as defined in \eqref{eq:PDP}, which better emphasizes spatial characteristics. Only the first 400 ns of data (40 points) are used, corresponding to a 10 ns resolution, representing a maximum path length of 120 m---sufficient for our analysis. Temporal features are extracted using 20 continuous measurements over a 20-second period, forming a PDP matrix $x \in \mathbb{R}^{40 \times 20}$, which serves as the RainGaugeNet input.
	
	RainGaugeNet extracts spatial features from individual measurements and then assembles these into a matrix for temporal feature extraction. The dual-path architecture ensures comprehensive spatiotemporal analysis, with RainGaugeNet(a) focusing on spatial features and RainGaugeNet(b) addressing temporal correlations. Both pathways utilize \textbf{ResNet1D} blocks, built from basic residual blocks (Fig.~\ref{fig:net}(c)).
	
	{\bf ResNet1D Structure}---Let the input be denoted by $x$ and the output by $y$. The residual block has two branches:
	\begin{itemize}
		\item The main branch, which transforms $x$ into $f(x)$ via two convolutional operations.
		\item The shortcut branch, which allows $x$ to bypass the transformations, contributing directly to the output.
	\end{itemize}
	This structure results in an output $f(x) + x$, mitigating the vanishing gradient problem and enhancing feature propagation.
	
	In the main branch, $x$ passes through two 1D convolutional layers with a kernel size of $3 \times 1$ (Conv 3x1). The first convolutional layer adjusts the number of channels, while the second maintains them. Each convolution is followed by batch normalization (BN) and a ReLU activation for the first convolution. In the shortcut branch,
	$x$ may undergo a $1 \times 1$ (Conv 1x1) convolution (Conv $1 \times 1$) with BN to ensure the dimensions of both branches match for seamless element-wise addition.
	
	{\bf Spatial Feature Extraction}---The PDP matrix has dimensions $\mathbb{R}^{\mathrm{batchsize} \times 40 \times 20}$. Initially, features are extracted from each time snapshot ($1 \times 40$), as shown in Fig.~\ref{fig:net}(a). A 1D convolutional layer with kernel size $5 \times 1$ (Conv $5 \times 1$, 128)captures these features, ensuring rain-induced multipath effects are preserved. BN normalizes the input distribution, stabilizing and accelerating training, while ReLU introduces non-linearity for learning complex multipath interactions.
	
	Subsequently, three ResNet1D blocks refine the extracted features, with output channels of 128, 256, and 64, respectively. These residual blocks enable hierarchical feature extraction, alleviating network degradation and improving model performance. The output is a matrix of dimensions $64 \times 6$, flattened and passed through two fully connected (FC) layers with output dimensions of 64 and 32, respectively. A ReLU activation follows the first FC layer, producing a 32-dimensional feature vector representing the input PDP.

	{\bf Temporal Feature Extraction}---After processing all 20 snapshots, the resulting matrix $32 \times 20$ is used to extract temporal correlations (Fig.~\ref{fig:net}(b)). Each individual PDP feature is treated as a separate channel. Temporal features are captured using a 1D convolution followed by consecutive ResNet1D blocks. The final representation is processed through three FC layers with output dimensions of 64, 16, and 3, providing the rainfall intensity classification. ReLU activation is applied after the first two FC layers, while the final layer excludes ReLU.

	In summary, RainGaugeNet employs multiple ResNet1D layers to fully extract multipath spatial features from a single snapshot and subsequently captures temporal correlations via additional ResNet1D layers. This dual-path design enables fine-grained processing, ensuring stable and efficient feature extraction even in deep architectures. The integration of residual connections prevents network degradation, enhancing the network's ability to classify rainfall intensity accurately based on 20 seconds of CSI data.
	
	\subsection{Data Collection and Training Parameters}

	\begin{figure}
		\centering
		\includegraphics[width=0.5\textwidth, trim=2cm 0cm 2cm 0cm, clip]{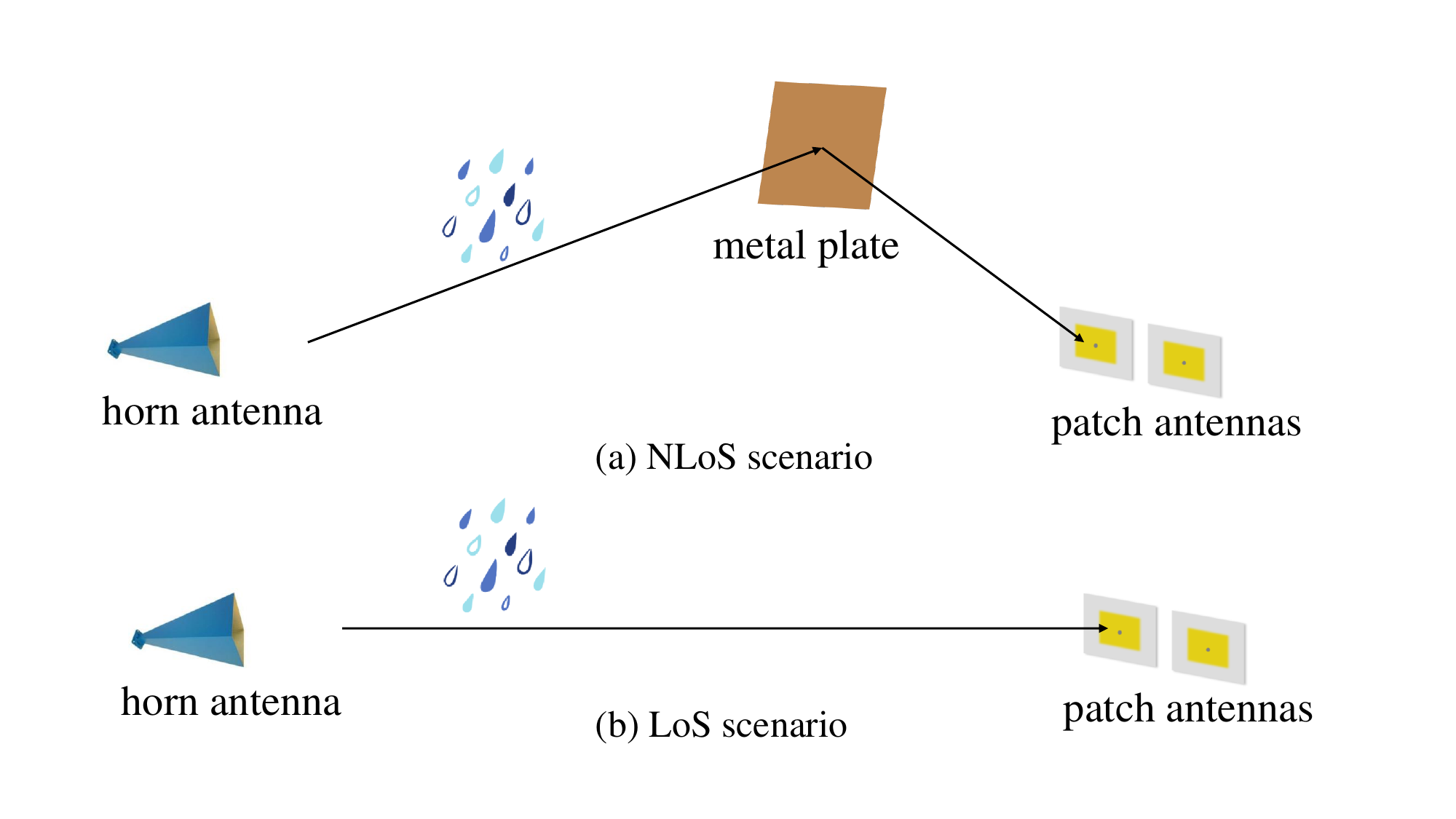}
		\caption{
			Illustration of the propagation scenarios:
			(a) NLoS scenario: the transmitted signal passes through a rainfall region and then undergoes reflection from a metal plate before reaching the patch antennas.
			(b) LoS scenario: the transmitted signal propagates directly to the patch antennas without reflection.
		}
		\label{fig:propagation}
	\end{figure}
	
	\begin{table*}
		\centering
		\normalsize
		\caption{Number of Samples in Training and Testing Sets}
		\label{tab:datanum}
		\begin{tabular}{llcccc}
			\toprule
			\textbf{Data Type} & \textbf{Condition} & \textbf{High wind (LoS)} & \textbf{Low wind (LoS)} & \textbf{High wind (NLoS)} & \textbf{Low wind (NLoS)} \\
			\midrule
			\multirow{3}{*}{Training Set}
			& No Rain       & 4410 & 5194 & 4410 & 3430 \\
			& Moderate Rain & 4410 & 8722 & 4410 & 3430 \\
			& Heavy Rain    & 4410 & 10486 & 4410 & 3430 \\
			\midrule 
			\multirow{3}{*}{Testing Set}
			& No Rain       & 630 & 700 & 630 & 490 \\
			& Moderate Rain & 630 & 700 & 630 & 490 \\
			& Heavy Rain    & 630 & 700 & 630 & 490 \\
			\bottomrule
		\end{tabular}
	\end{table*} 
	
	\begin{table*}
		\centering
		\normalsize
		\caption{Rainfall Classification Accuracy Results of Different Methods Under Various Conditions}
		\label{tab:result}
		\begin{tabular}{clcccc}
			\toprule
			\textbf{Condition} & \textbf{Weather Conditions} & \textbf{RainGaugeNet} & \textbf{RSS-Net} & \textbf{CNN} & \textbf{RainGaugeNet-single} \\
			\midrule
			\multirow{8}{*}{\shortstack{LoS}} 
			& Low wind speed - No Rain              & \textbf{100.00}   & 68.33   & 83.31   & 0.00   \\
			& Low wind speed - Moderate Rain        & \textbf{100.00}   & 25.62   & 97.85   & 27.89  \\
			& Low wind speed - Heavy Rain           & 80.14    & 61.91   & 79.69   & \textbf{100.00} \\
			& Low wind speed - Average              & \textbf{93.38}    & 51.95   & 86.95   & 42.63  \\
			\cmidrule(lr){2-6} 
			& High wind speed - No Rain               & \textbf{100.00}   & 54.49   & 100.00  & 0.00   \\
			& High wind speed - Moderate Rain         & \textbf{100.00}   & 13.73   & 99.91   & 0.00   \\
			& High wind speed - Heavy Rain            & \textbf{100.00}   & 77.90   & 62.41   & 100.00 \\
			& High wind speed - Average               & \textbf{100.00}   & 48.71   & 87.44   & 33.33  \\
			\midrule
			\multirow{8}{*}{\shortstack{NLoS}}
			& Low wind speed - No Rain              & \textbf{65.07}    & 19.36   & 54.29   & 0.00   \\
			& Low wind speed - Moderate Rain        & \textbf{100.00}   & 31.62   & 87.38   & 0.00   \\
			& Low wind speed - Heavy Rain           & \textbf{92.89}    & 78.55   & 71.57   & 81.37  \\
			& Low wind speed - Average              & \textbf{85.99}    & 43.18   & 71.08   & 27.12  \\
			\cmidrule(lr){2-6}
			& High wind speed - No Rain               & \textbf{100.00}   & 19.01   & 99.65   & 2.02   \\
			& High wind speed - Moderate Rain         & 58.10    & 70.77   & \textbf{72.54}   & 1.14   \\
			& High wind speed - Heavy Rain            & \textbf{100.00}   & 84.07   & 99.30   & 100.00 \\
			& High wind speed - Average               & 86.03    & 57.95   & \textbf{90.50}   & 34.39  \\
			\bottomrule
		\end{tabular}
	\end{table*}
	
	To ensure a comprehensive evaluation, we conducted data collection under two distinct weather conditions and two different signal propagation scenarios. Specifically, data were gathered at wind speeds below 5 m/s (low wind speed) and above 10 m/s (high wind speed). Additionally, measurements were taken in two propagation scenarios: a NLoS scenario involving reflections from a metal plate and a LoS scenario, as illustrated in Fig.~\ref{fig:propagation}. In total, data were collected under four experimental conditions.
	
	To minimize temporal correlation, the training and testing datasets were not randomly sampled. Random sampling could lead to overlapping or temporally adjacent data points between the training and testing sets, potentially introducing bias. Instead, data collection for the training and testing sets was conducted during temporally distinct periods, ensuring that each dataset was gathered in separate consecutive time intervals. Table~\ref{tab:datanum} provides detailed information about the training and testing datasets.
	
	The hyperparameters, including the learning rate, batch size, and training epochs, were set to 0.002, 512, and 30, respectively. The Adam optimizer was used with a weight decay of 0.1 to implement L2 regularization, effectively reducing overfitting. Additionally, a MultiStepLR scheduler was applied to adjust the learning rate, reducing it by a factor of 0.5 every 10 epochs. This strategy ensured effective convergence throughout the training process.

	\subsection{Baseline Algorithms}
	To ensure a fair comparison, we established three baseline rainfall classification networks:

	{\bf RSS-Net}---The first baseline model, RSS-Net, is based on RSS data and captures the temporal characteristics of RSS over consecutive time snapshots. This model serves to highlight the significance of the PDP matrix when compared with RainGaugeNet. The architecture consists of two one-dimensional convolutional layers: the first layer has 16 filters with a kernel size of 3, followed by a second layer with 32 filters. Max pooling is applied after each convolutional layer to reduce feature dimensions. The extracted features are flattened and processed through two fully connected layers. The first fully connected layer outputs 128 features with a ReLU activation, while the final layer provides the classification results.
	
	{\bf CNN}---The second baseline model utilizes the PDP matrix ($\mathbb{R}^{40 \times 20}$) as direct input to a two-dimensional convolutional neural network (CNN), illustrating the distinct roles of temporal and spatial feature extraction. In contrast, RainGaugeNet integrates these aspects more thoroughly, achieving fine-grained extraction.
	
	The CNN architecture begins with two 2D convolutional layers: the first supports 1 input channel and 16 output channels with a $3 \times 3$  kernel, stride of 1, and padding of 1 to maintain spatial dimensions. The second convolutional layer has 32 filters with identical specifications. Each convolutional layer is followed by a ReLU activation function. Max pooling layers with a kernel size of $2 \times 2$ are applied after each convolutional layer to downsample features, reducing computational complexity. The resulting feature maps are flattened and passed through two fully connected layers. The first fully connected layer outputs 128 features with a ReLU activation, while the final layer provides the classification result across 3 categories.
	
	{\bf RainGaugeNet-Single}---The third baseline model, RainGaugeNet-single, is designed to demonstrate the benefit of temporal-spatial feature fusion in RainGaugeNet. It shares the same architecture as RainGaugeNet but uses data only from the first time snapshot, with all subsequent snapshots set to zero. This setup directly evaluates the effect of integrating temporal features across multiple snapshots compared to relying solely on a single snapshot.

	These baseline models form a comprehensive framework for comparison, assessing the contributions of different aspects of the RainGaugeNet architecture. Specifically:
	\begin{itemize}
		\item {\bf RSS-Net} evaluates the use of RSS versus PDP.
		\item {\bf CNN} assesses the effectiveness of direct spatial feature extraction versus fine-grained temporal and spatial extraction.
		\item {\bf RainGaugeNet-Single} highlights the advantages of temporal-spatial feature fusion.
	\end{itemize}
	
	To ensure a fair and unbiased comparison, the training strategy for all three baseline models and RainGaugeNet is kept identical. The classification results of the various methods under different conditions are presented in Table~\ref{tab:result}.

	\subsection{Experimental Results and Analysis}
	
	RainGaugeNet demonstrates strong robustness, achieving high classification accuracy across all four experimental conditions. In the LoS scenario, which represents an ideal signal propagation environment, the PDP remains largely unaffected, allowing RainGaugeNet to achieve nearly 100\% accuracy in rainfall classification. However, under conditions of low wind speed with heavy rain, the accuracy drops to 80\%. While lower, this is still significantly superior to other baseline models. Notably, the RainGaugeNet-single model, which tends to classify almost all conditions as heavy rain, achieves 100\% accuracy in heavy rain scenarios due to this bias.
	
	In contrast, RSS-Net performs poorly, with accuracy sometimes dropping below 20\%. RainGaugeNet, leveraging CSI, shows a significant improvement, consistently outperforming RSS-based methods by approximately 40\%. {\bf This underscores the capability of CSI to provide detailed channel state information, enhancing discrimination between different rainfall intensities.}
	
	The inferior performance of CNN compared to RainGaugeNet can be attributed to its limited granularity in capturing temporal-spatial features. RainGaugeNet's approach of separately treating PDP characteristics and temporal features allows for more fine-grained and physically meaningful feature extraction. Conversely, the CNN processes these features simultaneously, resulting in less effective classification. While RainGaugeNet-single, relying solely on PDP data from a single time snapshot, fails to provide meaningful rainfall classification, the CNN---despite not matching RainGaugeNet's performance---shows better results by successfully identifying most rainfall conditions using the same PDP data. Both CNN and RainGaugeNet outperform RainGaugeNet-single, {\bf demonstrating the importance of incorporating temporal correlation features for accurate rainfall classification.}
	
	In the NLoS scenario, where signals undergo reflection off a metal plate after passing through rain, overall classification performance is lower compared to the LoS environment. All networks experience a decline in accuracy under NLoS conditions. For medium rainfall with both low and high wind speeds, classification accuracy drops to approximately 60\%. Nevertheless, RainGaugeNet maintains nearly 100\% accuracy for other conditions, highlighting its robustness. RSS-Net and RainGaugeNet-single fail to provide usable results in these scenarios, while the CNN model also exhibits degraded performance, still lagging behind RainGaugeNet. This decline is likely due to signal reflections from the metal plate, which significantly alter the multipath environment and change the PDP distribution, making accurate classification more challenging.
	
	{\bf On average, RainGaugeNet achieves over 90\% classification accuracy in LoS environments. When the PDP distribution is altered by reflections in NLoS conditions, the accuracy drops slightly but remains above 85\%, demonstrating sufficient stability in rainfall classification.}
	
	Our experimental investigations into the proposed CSI-based RainGaugeNet provide strong evidence of its effectiveness in rainfall classification across various conditions.

	\section{Conclusion}
	\label{sec:conclusion}
	This study introduced the first CSI-based sub-6 GHz rainfall attenuation measurement and analysis. Utilizing a 2.8 GHz artificial rainfall channel measurement system, we investigated the impact of varying rainfall intensities using CSI as a detailed metric. Our measurements under no rain, moderate rain, and heavy rain conditions revealed that energy variance is highest in no rain scenarios, aligning with previous findings, while the differences between moderate and heavy rainfall variances were minimal.
	Multipath analysis showed that delay spread increased under rainfall but slightly decreased with higher rainfall intensity. The attenuation channel followed a power-law decay model, with decay rates decreasing as rainfall intensified.
	To classify different rainfall intensities, we proposed RainGaugeNet, the first CSI-based rainfall classification model capable of extracting both multipath and temporal features from consecutive CSI snapshots. RainGaugeNet achieved over 90\% classification accuracy in LoS scenarios and over 85\% in NLoS scenarios, demonstrating its robustness and effectiveness across various conditions.
	
	\bibliographystyle{IEEEtran}
	\bibliography{paper}	
\end{document}